\documentclass{vgtc}                          

\usepackage{times}                     

\usepackage{mathptmx}                  
\usepackage{amsmath,amssymb}

\usepackage{tabularx}
\usepackage{subcaption}
\usepackage{multirow}
\usepackage{multicol}
\usepackage{float}

\onlineid{0}

\vgtccategory{Research}

\vgtcinsertpkg

\title{Statistical Blendshape Calculation and Analysis for Graphics Applications}

 \author{Shuxian Li\thanks{e-mail: lisx14@lenovo.com}\\ %
         \scriptsize Lenovo Research %
 \and Tianyue Wang\\ %
      \scriptsize Lenovo Research %
 \and Chris Twombly\thanks{e-mail: ct593@cornell.edu}\\ %
      \scriptsize }

\teaser{
  \centering
  \includegraphics[width=\linewidth]{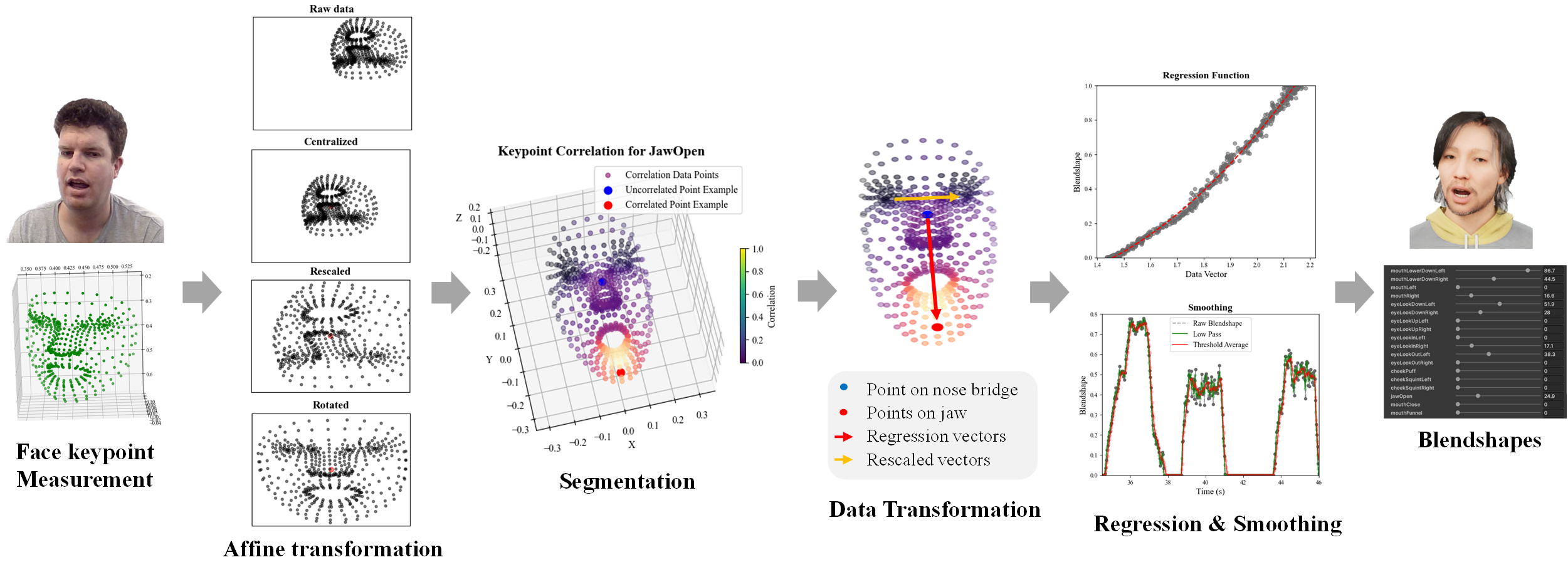}
  \caption{Logical process flow of blendshape estimation method}
  \label{fig_1}
}

\abstract{ With the development of virtualization and AI, real-time facial avatar animation is widely used in entertainment, office, business and other fields. Against this background, blendshapes have become a common industry animation solution because of their relative simplicity and ease of interpretation. Aiming for real-time performance and low computing resource dependence, we independently developed an accurate blendshape prediction system for low-power VR applications using a standard webcam. First, blendshape feature vectors are extracted through affine transformation and segmentation. Through further transformation and regression analysis, we were able to identify models for most blendshapes with significant predictive power. Post-processing was used to further improve response stability, including smoothing filtering and nonlinear transformations to minimize error. Experiments showed the system achieved accuracy similar to ARKit 6. Our model has low sensor/hardware requirements and realtime response with a consistent, accurate and smooth visual experience.
} 


\begin{document}



\firstsection{Introduction}

\maketitle

With the rapid development of AI and VR fields, digital humans, as an important component, have a wide range of potential application in the fields of social media, business, and entertainment. Many key technologies are used with virtual humans including avatar creation, animation, rigging, interaction, application management. Among them, facial capture and driving technology are some of the most important components needed for realization of virtual digital human applications. These technologies allow users to dynamically control expression of virtual avatars in real time. 

Human avatars are often represented using a virtual mesh model. The 3D mesh vertex positions need to be fully specified so graphics programs can render the character. The main types of character animation include skinned mesh animation, joint animation and morph animation. Storing all mesh vertex positions for each video frame is very accurate, but requires significant computing resources and is not suitable for real-time scenarios. Joint animation is sometimes used to control eye movement, which is simple and convenient, but it is not suitable for fine facial expression control. Morphable mesh animation absorbs their advantages, limiting the computational cost while maintaining high accuracy 

The morphable model method determines changes in the character mesh model using small deformations controlled by dynamic parameters. Among the many deformable methods, blendshapes are popular due to relative computational simplicity, physical interpretability and convenience. Animators provide the target shape and then interpolate between the neutral shape mesh and the target expression mesh to generate animations. This method is widely used to animate dynamic expressions for virtual characters. The method has real-time performance and can accurately map face states using specific expression definitions. 

The most popular methods for avatar animation include full facial reconstruction, pure image transform approaches, landmark-driven, and audio-driven techniques. Most of these methods require large face datasets for training, and the high computational resource requirements that limit their application for low-performance devices or for real time animation. Blendshapes provide an solution to these problems, this work proposes a real-time facial animation method using a webcam. We transform detected landmarks using a statistical model into blendshape coefficients used for 3D character animation. The independently developed model has higher accuracy, lower hardware requirements, relies on less training data, and is suitable for low-power computing scenarios. Our contributions are:

\begin{itemize}
    \setlength\itemsep{0em}
    \item Independently developed statistical models to do real time conversion from landmarks to blendshapes using a webcam.
    \item  A set of algorithms designed for optimizing animation results.
    \item  Extensive statistical testing, including accuracy comparison and hardware utilization results that show near state-of-the art performance with lower hardware requirements.
\end{itemize}

\section{Background}


\subsection{Mesh to Blendshape}

Mesh models are a common way to represent 3D faces and is widely used in video games, animation, engineering, design and other fields. A mesh consists of vertices, edges and faces, which can efficiently represent complex 3D shapes. Each vertex has coordinates in a 3D virtual space. The surface of a 3D object can be accurately described and rendered through these vertices and their connected edges and faces, and by moving the positions of these vertices, different facial expressions can be represented.

Delicate and accurate facial expression modeling requires the detailed calculation of all mesh vertex locations along with required dynamic transformations to update, but full reconstruction often requires large computing resources. Three dimensional morphable models (3DMM) simplifies the modeling process by reducing computational load. In 1999 A Morphable Model for the Synthesis of 3D Faces\cite{blanz_morphable_1999} widely popularized 3DMM human models for face reconstruction and character animation. The authors then used principal component analysis to identify common features for human faces. The identified features were fitted to a multidimensional normal distribution, and then used statistical distance from the fitted distribution to define a set of fitting parameters and basis vectors. They were able to use linear combinations of these feature vectors to build new human faces. Additional linear and nonlinear face models and datasets have been developed, including Basel\cite{paysan_3d_2009,gerig_morphable_2017,gerig_morphable_2018}, Surrey\cite{huber_multiresolution_2016}, Facewarehouse\cite{cao_facewarehouse_2014}, large scale face model\cite{booth_large_2018}, and 4D face model\cite{wu_mmface4d_2023}. All of these methods independently developed dynamic parameters that can be used to generalize faces to match specific features.

Blendshapes are related to 3DMM, expanding on the common feature idea to describe dynamic face movements, with a simple interpretation and clearer definition. Blendshapes are a set of pre-defined face states that form a semantic parameterization of face movement: 
\begin{equation}
f= w_o f_o+w_1 f_1+...
\label{eq_1}
\end{equation}
\noindent Where $w_o,...$ are the weight coefficients, and $f_o,...$ are a set of face expression states. Weight coefficients range between 0 to 1. Each term represents a specific sub-expression used to define a feature set of an expression space. 

Blendshape models are popular tools for VR animation because of their relative simplicity, clear physical definition/interpretability, and ability to restrict expression to highly plausible ranges of motion. They have become the primary method of animation for the movie\cite{lefebvre_practice_2014} and videogame\cite{jandric_rendering_2023} industries. Blendshape like feature control has been used in character animation since at least the 1970s\cite{parke_fredric_i_parametric_1974}, and implemented in commercial software by the late 1980s\cite{lewis_reducing_2005}. Recent effort have focused on using computer vision and AI approaches to estimate blendshape states for human faces, and then use blendshapes to animate human avatars.

Blendshapes have traditionally been individually created for each human model by the animator. The number of blendshapes for characters varies between 50 to 900+ depending on the accuracy and desired expression range\cite{lefebvre_practice_2014}. For most video game applications a set of 52 blendshapes has become a de-facto common standard, used by the most common graphics engines for game development including Unreal Engine 5\cite{epic_games_unreal_2023} and Unity\cite{technologies_unity_nodate}. This is based on the Apple augmented reality kit (ARKit\cite{noauthor_arkit_nodate}). ARKit blendshapes were created using PCA decomposition and cluster analysis to describe common expression patterns. ARKit blendshapes use the high-resolution camera/lidar/IR sensor systems found in an iPhone 10 or later devices. This does not match less powerful systems like standard webcams. This work extends blendshapes to webcam based AI landmark regression models. We perform direct comparative analysis between our algorithm and ARKit 6.

\subsection{Landmark Regression}


Facial key point detection methods can be roughly divided into the following categories: traditional methods such as ASM \cite{ASM_1995} and AAM \cite{AAM_1998}; methods based on cascade shape regression, such as CPR\cite{CPR_2010}; methods based on deep learning, such as DCNN\cite{DCNN_2013}, MTCNN\cite{MTCCN_2016} and DAN\cite{DAN_2017}. There are many full body pose estimation algorithms, including facial landmark detection solutions, such as RTMPose\cite{RTMPose_2023}, DWPose\cite{DWPose_2023} and OpenPose\cite{openpose_2019}. Our algorithm uses the output from a landmark detection, specifically a method with high real-time performance and sufficient state position information to identify clear movements. Most of the above methods lack real-time performance or lack sufficient facial landmarks. The Google MediaPipe Holistic model\cite{lugaresi_mediapipe_2019} meets these requirements for a facial geometry solution that can estimate 478 3D face/eye landmarks in real time with only a monocular camera input.

Mediapipe uses a two step calculation process to face detection. The model first performs face segmentation using BlazeFace\cite{bazarevsky_blazeface_2019}, which is based on MobileNet\cite{howard_mobilenets_2017,sandler_mobilenetv2_2018} but further optimized for speed. The net effect of these optimizations results in a 3.5x reduction in inference time compared to MobileNetV2, but with a 40\% increase in regression error and a 47\% increase in jitter\cite{bazarevsky_blazeface_2019}. The jitter is especially noticeable for sensitive blendshape calculation, and required the implementation of smoothing algorithms to compensate for this effect. Following face detection and segmentation, a second process using a deep learning model\cite{kartynnik_real-time_2019} for landmark regression using Procrustes Analysis\cite{gower_generalized_1975} to map the regressed landmarks onto a 3D metric space. The returned z-coordinates represent depth relative to a reference plane passing through the mesh’s center of mass\cite{kartynnik_real-time_2019}. Compared to other landmark regression methods MediaPipe detects a higher number of facial landmarks at real time inference speeds on low power computing devices. 

This research relied on the MediaPipe Holistic Model 0.9.0.1\cite{lugaresi_mediapipe_2019}, but can be extended for other implementations. Considering about the limitations of mediapipe, more work is needed to address limitations, including rotation, anchoring, and some nonlinear face distortion.

\section{Methods}

Our model converts face landmarks to blendshapes in real time using a standard webcam. To do this the model performs the following tasks:

\begin{equation}
    f_b=F(R(T_d (S(T_a (X)))))
    \label{eq_2}
\end{equation}

\noindent Where $X$ represents the regressed facial landmarks and $f_b$ are the blendshape estimates. The equation terms are:
\begin{itemize}
    \setlength\itemsep{-0.5em}
    \item $T_a$: Affine Transformation
    \item $S$: Segmentation 
    \item $T_d$: Data Transformation
    \item $R$: Regression
    \item $F$: Smoothing 
\end{itemize}

\noindent Each term in the equation is supported with statistical testing. The calculation process is demonstrated in \cref{fig_1}.

\subsection{Affine Transformation}

\begin{figure}[h]
    \centering
    \includegraphics[width=\linewidth]{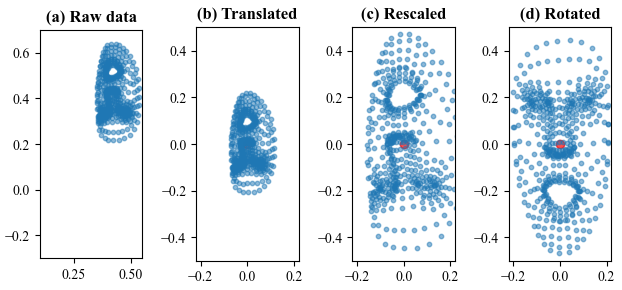}
    \caption{Effect of the affine transformation matrices on (a) Raw data where (b) $R_1$, (c) $R_2$ (d) $R_3$ are applied in succession}
    \label{fig_aff}
\end{figure}

To minimize regression error, we perform a series of affine transformations. Affine transformations maintain collinearity between point sets, parallelism between lines, Convexity between sets, ratios of vector magnitudes and geometric centroids. Transformations are chosen so that landamrks are displayed in a common, relatively rotation and distortion invariant basis. This is suitable for further data transformation and regression. The complete affine transform is:

\begin{equation}
    T_{a}= R_{3} \cdot R_{2} \cdot R_{1} \cdot X
    \label{eq_3}
\end{equation}

\noindent Where $X$ are raw landmark position estimates, $R_1$, $R_2$, $R_3$ are transformation matrices used to map the observed face to the desired regression basis, and $T_a$ is the transformed output. Each matrix represents a specific set of rotation, transformation, and resealing operations to transform the input landmarks into an eye-eye-nose screen facing basis for regression, shown in \cref{fig_aff}.

\subsection{Segmentation}

The next step in blendshape processing is landmark segmentation. Segmentation is useful because it can reduce the dimensionality of data while maintaining high accuracy using statistical analysis to identify specific relevant features. This has the effect of reducing memory and computation requirements by 95-99\%. Based on the relative sparsity of the training data, correlation analysis, feature regression, and support vector regression were used for feature identification. This was sufficient to identify features for most blendshapes. One Notable exception is TongueOut. This makes intuitive sense, since no MediaPipe Holistic keypoints are used to describe tongue movement. 

\begin{figure}[h]
    \centering
    \includegraphics[width=\columnwidth]{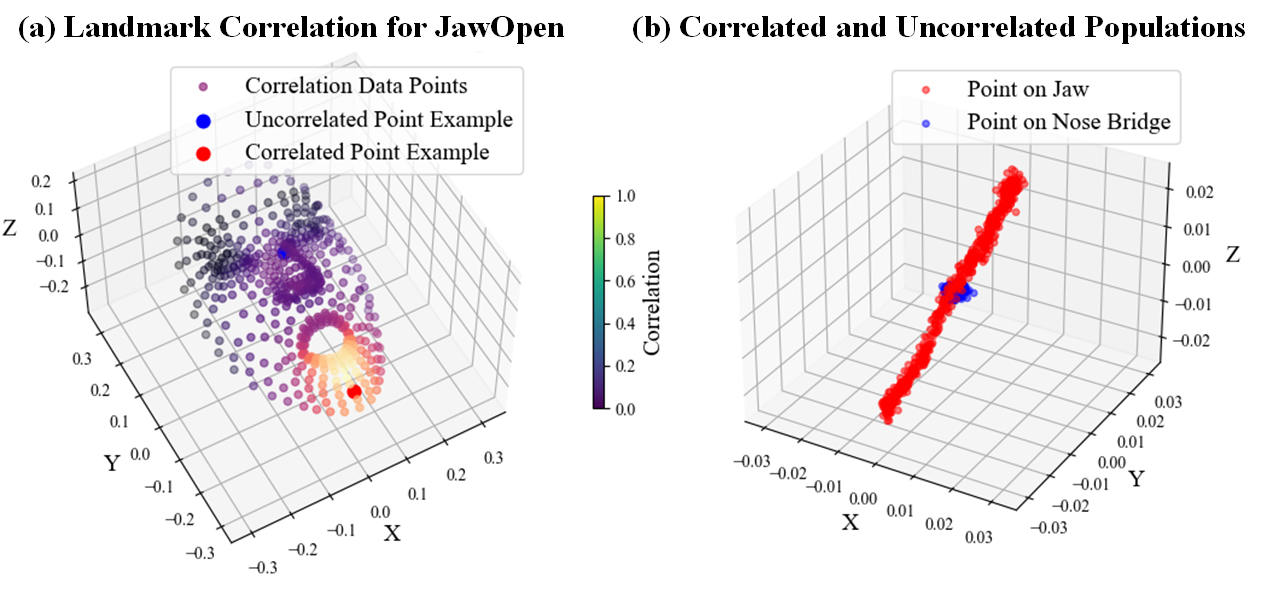}
    \caption{Correlation for JawOpen. (a) Linear correlation magnitude for all points on the face for JawOpen. Yellow points are highly correlated, while black is uncorrelated. (b) Behavior of highly correlated (blue) and uncorrelated (red) sample points. This comparison demonstrates distinct behavior that justifies segmentation before regression modeling.}
    \label{fig_seg}
\end{figure}

\begin{figure}[h]
    \centering
    \includegraphics[width=\columnwidth]{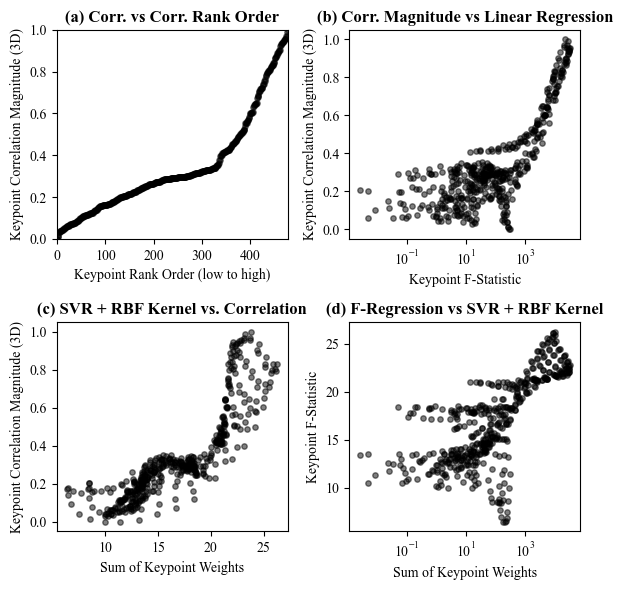}
    \caption{Statistical support for keypoint selection and segmentation. (a)-(d) Shows further statistical analysis. (a) Landmark rank order, (b) 3D correlation vs. 1D linear regression, (c) SVR + RBF kernel vs 3D correlation, (d) F-regression vs. SVR + RBF.}
    \label{fig_seg2}
\end{figure}

More comprehensive analysis is demonstrated with \cref{fig_seg} with JawOpen. We observed distinct behavior differences between highly correlated and uncorrelated landmark points using synthetic test data, shown in \cref{fig_seg}(a)-(b). Correlation analysis identified the top 2 percentile of highly correlated keypoints. Several cross-validation methods were used to check the relative importance of the selected landmarks, including F-regression and a Support Vector Regressor with a radial basis kernel. We found good agreement between different landmark selection methods, as demonstrated with \cref{fig_seg2}(a)-(d).

Based on this analysis, segmentation matrices were developed for each blendshape model. This operation is represented using a set of resampling matrices used to select specific keypoint subsets:

\begin{equation}
S=D \cdot T_a
\label{eq_4}
\end{equation}

\noindent Where $D$ is the segmentation matrix, $T_a$ are the transformed landmark estimates, and S is the segmented output. We used a eye-nose centered basis for our statistical modeling.

\subsection{Transformation}

Data transformations are necessary to make feature data more suitable for regression, especially when the required linear relationship between the independent and dependent variables is difficult to observe. It is sometimes possible to transform the independent variables to restore linearity, and therefore suitability for standard regression. Data transformation can result in reduced prediction error, and can be motivated by statistical testing. Common requirements of standard regression methods include homoscedasticity (homogeneity of variance) and multivariate normality. We tailored the final data transformation step for each model based on individual model requirements and observed statistical properties of the landmark data.

\begin{figure*}[!t]
    \centering
    \begin{subfigure}{\textwidth}
        \includegraphics[width=\linewidth]{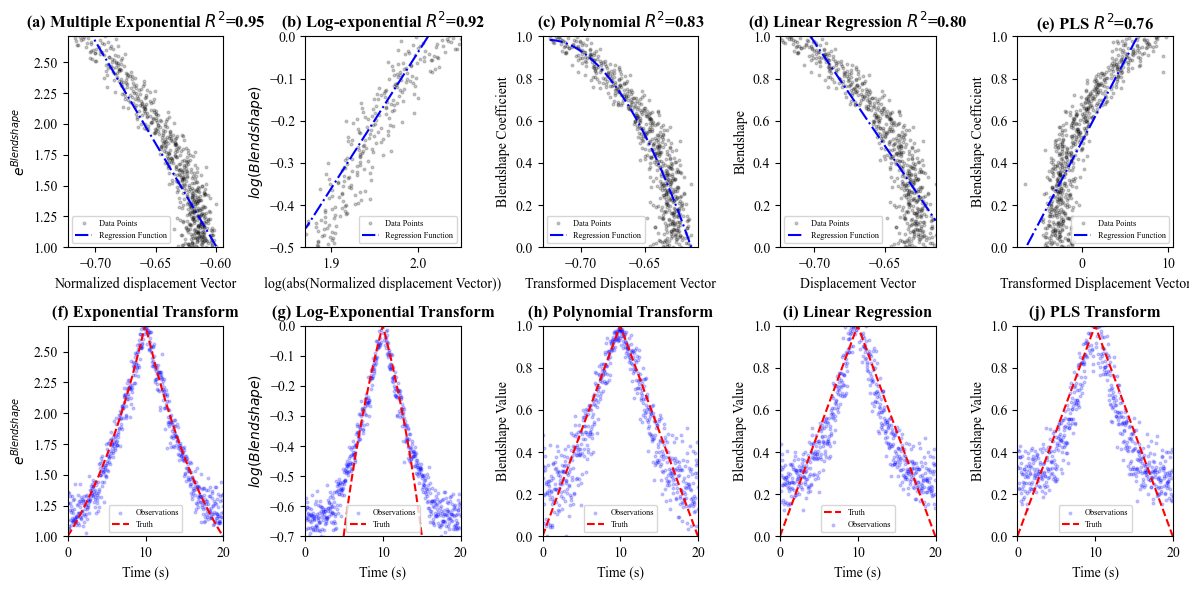}
    \end{subfigure}
    \begin{subfigure}{\textwidth}
        \centering
        \renewcommand\tabularxcolumn[1]{m{#1}}
        \renewcommand\arraystretch{1.3}
        \setlength\tabcolsep{2pt}
    \begin{tabularx}{\linewidth}{*{6}{>{\centering\arraybackslash}X}}
    \hline
\textbf{Test/Metric} & \textbf{Exponential} &\textbf{Log-Exponential} & \textbf{Polynomial} & \textbf{Linear} & \textbf{PLS} \\
    MSE\cite{johnson_applied_2002} & 0.0119 & 0.016 & 0.0134 & 0.0163 & 0.0196\\
    Durbin-Watson\cite{johnson_applied_2002} & 1.360 & 1.261 & 1.140 & 0.656 & 0.680\\
    Breusch-Pagan\cite{breusch_simple_1979} & 21.5 p=\(2.41\cdot 10^{-4}\) & 13.4 p=\(9.15\cdot 10^{-3}\) & 49.1 p=\(2.12\cdot 10^{-11}\)& 17.6 p=\(2.68\cdot 10^{-5}\) & 13.3 p=\(2.57\cdot 10^{-4}\)\\
    F-Statistic\cite{weir_estimating_1984} & 2856 & 675.3 & 1556 & 2463 & 1946\\
    Cor(res.-tr.)\cite{pearson_liii_1901} & 0.975 & 0.967  &  0.915  &  0.897 & 0.874 \\
    \(\xi\)\cite{chatterjee_new_2021}& 0.778 & 0.679 & 0.658 & 0.658 & 0.623 \\
    \hline
    \end{tabularx}
    \end{subfigure}
    \caption{Data transformation and sample regression for CheekPuff, and resulting coefficients of determination (\(R^2\)). \textbf{(a)} Shows a multiple exponential model, \textbf{(b)} multiple log-exponential model, \textbf{(c)} polynomial (deg=2) , \textbf{(d)} Standard OLS Linear Regression \textbf{(e)} Partial Least Squares projection (PLS, deg=1), \textbf{(e)}. Beneath each is a truth comparison for \textbf{(f)} exponential \textbf{(g)} log-exponential, \textbf{(h)} polynomial, \textbf{(i)} Linear \textbf{(j)} PLS. Statistical measures of performance include Mean squared error\cite{johnson_applied_2002}, Durbin-Watson statistic\cite{durbin_testing_1950}, Breush-Pagan test\cite{breusch_simple_1979}, Fisher-Statistic\cite{weir_estimating_1984}, Pearson correlation\cite{pearson_x_1900} and \(\xi\) correlation\cite{chatterjee_new_2021}.}
    \label{fig_3}
\end{figure*}

As an example, for some blendshapes we used principal component decomposition\cite{hotelling_relations_1936,pearson_liii_1901} (PCD) as a transformation method. This transforms landmark displacement into a new coordinate system along the direction of greatest variance. Variance stabilization and other transformations could then applied to produce better resolved features for regression. We used transformations that act on either the landmark estimates:
\begin{equation}
  T_d = T(S)   
  \label{eq_5}
\end{equation}
\noindent Where $T$ is a function applying a transformation to the segmented landmark estimates S, and $T_d$ is the result of the transform. Or on the distribution of dependent variables used to develop individual regression models:

\begin{equation}
\epsilon = T_{d}(Y - mx)
\label{eq_6}
\end{equation}

\noindent Where the transformation operation $T_d$ is applied to the distribution of training data used to create a regression model $mX + \epsilon$. Both types of transformations are used.

A demonstrative example is shown in \cref{fig_3}, which compares the effects of different transformation methods on the accuracy of ordinary least squares regression for CheekPuff. Graphs \cref{fig_3}(a)-(e) shows data fitting, and (f)-(j) shows an accuracy comparison to truth data. The table shows the results of different statistical tests to quantify model accuracy, including coefficient of determination ($R^2$)\cite{johnson_applied_2002}, Mean squared error\cite{johnson_applied_2002}, Durbin-Watson statistic\cite{durbin_testing_1950}, Breush-Pagan test\cite{breusch_simple_1979}, Fisher-Statistic\cite{weir_estimating_1984}, Pearson linear correlation between truth and response\cite{pearson_x_1900} and $\xi$ correlation\cite{chatterjee_new_2021}. The Durbin-Watson test detected heteroskedasticity, partly compensated by the use of the transformation operations.

\subsection{Regression}

Regression is a statistical process for estimating the relationship between a set of detection and response variables. Our blendshape estimation system uses several different regression models, depending on the relative significance and usefulness of detected regression features following the methods highlighted in \cref{fig_3}. Most blendshapes can be accurately estimated using relatively less complex methods, such as multiple linear ordinary least squares regression. This is justified using statistical analysis. The requirements for these models to obtain accurate predictions include multivariate normality and homoskedasticity. After final data transformation, approximately 10 blendshapes require more complex regression methods. This is highlighted with CheekPuff, which has significant heteroskedasticity that cannot be accounted for with simpler regression methods. 

\begin{figure}[h]
    \includegraphics[width=\columnwidth]{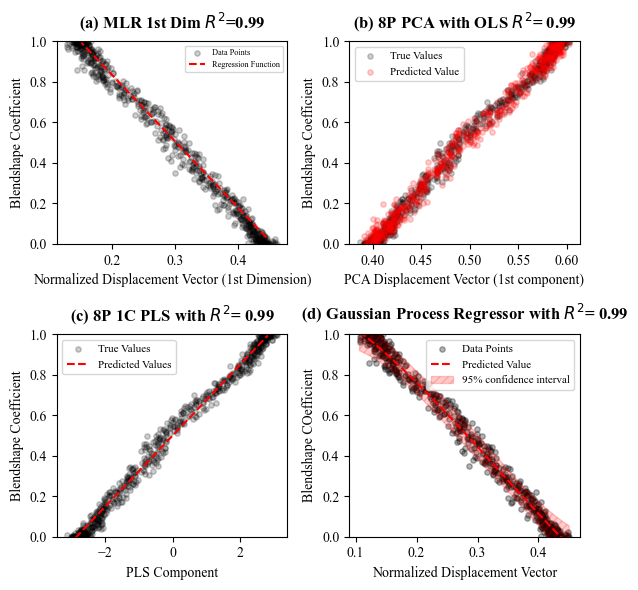}
    \caption{Regression models for simple blendshape classes. (a) Single element of a multiple linear regression model. (b) A 1 component PCA model using 8 different data vectors. (c) A PLS model created using 8 keypoint displacement vectors. (g) A Gaussian process regression model showing the 95\% confidence interval for the probability distribution.}
    \label{fig_easyreg}
\end{figure}

\begin{figure}
    \centering
    \includegraphics[width=\columnwidth]{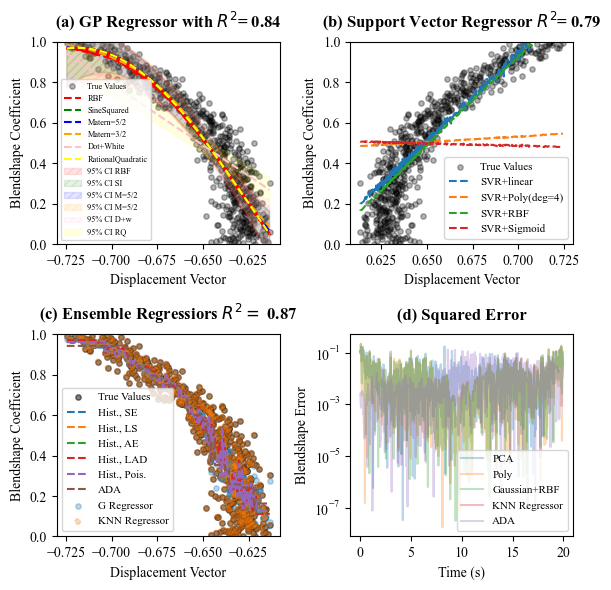}
    \caption{Non-standard regression models for heteroskedastic blendshape estimation. (a) Gaussian process regressors with different kernel functions. (b) Support vector regressors with different kernel functions. (c) Ensemble regressors with different kernel functions. (d) Squared error comparison of different regression models. Some blendshapes require more complex regression procedures.}
    \label{fig_reg2}
\end{figure}

Blendshapes that satisfy or nearly satisfy linear regression requirements can be successfully modeled with standard techniques. Examples include multiple linear regression, polynomial regression, PCA regression and PLS regression, demonstrated in \cref{fig_easyreg}. We tested more advanced regression methods including Gaussian process regression with different kernel functions, support vector regression with different kernel functions and different ensemble methods (histogram gradient boosted methods, ADA regression, gradient regression, and k-nearest neighbor regression), demonstrated in \cref{fig_reg2}. The model choice was justified using statistical testing, using $R^2$\cite{johnson_applied_2002}, $\xi$\cite{chatterjee_new_2021}, mean squared error\cite{johnson_applied_2002}, Durbin-Watson[38] test for autocorrelation\cite{durbin_testing_1950} (DW), residual error\cite{johnson_applied_2002}, Shapiro-Wilks test\cite{shapiro_analysis_1965} (SW) and memory utilization, shown with \cref{tab_reg}. 

\begin{table}
    \caption{Demonstrative statistical behavior of blendshape models}
    \resizebox{\columnwidth}{!}{%
    \centering
    \begin{tabular}{lcccccc}
    \hline
    \textbf{Model} & \textbf{\(R^2\)\cite{johnson_applied_2002}} &\textbf{\(\xi\)\cite{chatterjee_new_2021}} & \textbf{MSE\cite{johnson_applied_2002}} & \textbf{DW\cite{durbin_testing_1950}} & \textbf{SW\cite{shapiro_analysis_1965}} & \textbf{Memory (Bytes)} \\
    \hline
    \multicolumn{7}{c}{\textbf{Standard Blendshapes (SmileLeft)}} \\
    ML-OLS & 0.9919 & 0.8849 & 0.0010 & 1.7679 & 0.9976 & 1553\\
    Poly   & 0.9882 & 0.8648 & 0.0014 & 1.6839 & 0.9931 & 20856\\
    PCA    & 0.9918 & 0.8501 & 0.0018 & 1.7380 & 0.9964 & 4192\\
    PLS    & 0.9899 & 0.8659 & 0.0015 & 1.6906 & 0.9975 & 5358\\
    GP/RBF & 0.9886 & 0.8645 & 0.0014 & 1.6274 & 0.9931 & 2911867\\
    SVR/RBF & 0.9871 & 0.8680 & 0.0015 & 1.6768 & 0.9976 & 1602\\
    GBH/RBF & 0.9905 & 0.9112 & 0.0011 & 1.8433 & 0.9855 & 2707469\\
    \hline
    \multicolumn{7}{c}{\textbf{Complex Heteroskedastic (CheekPuff)}}\\
    ML-OLS & 0.9365 & 0.7812 & 0.0120 & 0.5719 & 0.9858 & 1548 \\
    Poly   & 0.8458 & 0.6577 & 0.0134 & 1.1404 & 0.9663 & 30488 \\
    PCA    & 0.7560 & 0.6235 & 0.0203 & 0.6800 & 0.9811 & 5486\\
    PLS    & 0.7649 & 0.6237 & 0.0196 & 0.6800 & 0.9795 & 14062\\
    ML-Exp & 0.9509 & 0.7784 & 0.0119 & 1.3601 & 0.9392 & 1548 \\
    GP/RBF & 0.8405 & 0.6587 & 0.0133 & 0.9960 & 0.9710 & 2897439\\
    SVR/Poly & 0.0873 & 0.6430 & 0.0763 & 0.0005 & 0.9539 & 55957 \\
    SVR/RBF & 0.8038 & 0.6605 & 0.0164 & 0.5063 & 0.96674 & 32721\\
    GBHist/SE & 0.8715 & 0.7597 & 0.0107 & 1.1232 & 0.9766 & 26837469 \\
    ADA & 0.8517 & 0.9310 & 0.0123 & 0.8351 & 0.9901 & 8028 \\
    Grad. Bost. & 0.9043 & 0.7943 & 0.0080 & 1.171 & 0.9788 & 5529 \\
    \hline
    \end{tabular}
    }
    \label{tab_reg}
\end{table}

Simpler regression methods are usually more robust to small observational error or edge case conditions. More complex regression methods make fewer assumptions about the underlying data distribution, but come with additional computational cost and loss of clear interpretation. We observe significant variations in performance between models, as well as the presence of strong autocorrelation for some blendshapes, including SmileLeft and CheekPuff. These regression models can accurately predict blendshape coefficients for specific face models. Additional complication comes when these models are applied to different faces.

This introduces significant bias, which motivated the inclusion of extra terms in the regression model:

\begin{equation}
    R_{t}(f)=(1+\frac{g_{t}(f)}{\eta_{t}} -\gamma_{t})(w\cdot f+b+\beta_{t})
    \label{eq_7}
\end{equation}

\noindent Where $wf + b$ describes the standard components of a linear regression model, the weighting coefficients and the initial bias, $g(f)$ represents a response function, $\eta$ is a damping term, $\gamma$ is a time weighted bias correction term, and $\beta$ is an additional bias correction term. The $\beta$ term is calculated by:

\begin{equation}
   \beta_{t,0} = \begin{cases}
        f_{t} \;\;\; f_{t} > f_{t_{o}} + e^{k \cdot (t_{o}-t)}\\
        \beta_{0} \;\;\; f_{t} \leq f_{t_{o}} + e^{k \cdot (t_{o}-t)}\\
    \end{cases}
    \label{eq_8}
\end{equation}

\noindent Where $k$ is an adjustable time scale, $F_t$ is the current blendshape estimate, $B_t$ is the shift term, and $t_t-t_o$ is the time difference between the previous minimum value and the current frame. The $\gamma$, $\eta$ terms can either be derived empirically or from statistical testing of landmark and regressed blendshapes, are customized for each blendshape.

\begin{figure}[h!]
    \centering
    \includegraphics[width=\columnwidth]{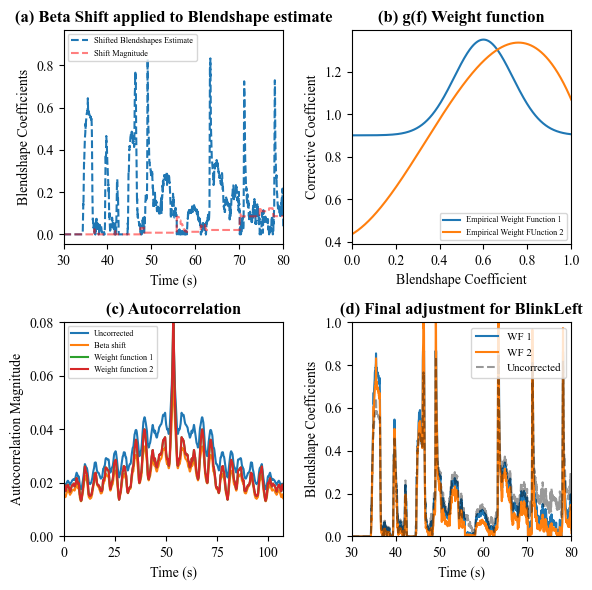}
    \caption{Bias correction and autocorrelation for BlinkLeft. (a) Shows the effect of $\beta$ shift on the raw blendshape estimates. (b) Two empirical weighting functions used for bias correction. (d) The beta shift has a large impact on autocorrelation between blendshape observations, following weighting function, damping, and shift terms do not change autocorrelative behavior. (d) Initial and corrected blendshape estimates during testing.}
    \label{fig_nonlinear}
\end{figure}

The effect of the nonlinear corrective terms is shown for BlinkLeft with \cref{fig_nonlinear}. For most blendshapes it is especially difficult to observe the last 15\% of dynamic range in real world faces. \cref{fig_nonlinear}(a) Demonstrates the effect of the $\beta$ term in equations \ref{eq_7} and \ref{eq_8}. This has a significant impact on blendshape autocorrelation shown in \cref{fig_nonlinear}(c). The greatest change in autocorrelation is caused by the \(\beta\) shift, while the following weighting function, damping, and shift terms do not noticeably change blendshape behavior. Two empirically derived weight functions are shown in \cref{fig_nonlinear}(b). These functions are designed to correct regression bias, especially near the upper edges of the blendshape range. \cref{fig_nonlinear}(d) shows the cumulative impact of all corrective terms on blendshape estimates shown during testing.

\subsection{Smoothing}

We observed significant landmark jitter that negatively impacted the stability of our blendshape estimates. This is especially evident under edge conditions such as partial obstruction or near full head rotation. This motivated development of smoothing algorithms compensate for these effects, while providing additional bias correction and faster response time than standard interpolation methods. There was significant difference in desired response behavior between blendshapes, so several different smoothing algorithms were implemented, including:

\begin{itemize}
    \setlength\itemsep{-0.5em}
	\item Kalman filter\cite{kalman_new_1960}
	\item Moving Average\cite{isufi_autoregressive_2017,harvey_maximum_1979}
	\item Time weighted moving average\cite{hunter_exponentially_1986}
    \item Low pass filtering
	\item Gated moving average
\end{itemize}

The gated moving average is designed to relax filtering when there is a sudden large change in response behavior. This has the benefit of reduced smoothing lag while retaining the smoothing power of averaging during normal response. This can be represented using the following equations:

\begin{equation}
f_{avg}= \frac{1}{N} \sum_{i=1}^{N} f_i
\label{eq_9}
\end{equation}

\begin{equation}
f_{var}= \sqrt{\frac{1}{N-1} \sum_{i=1}^{N}(f_i - f_{avg})^2}
\label{eq_10}
\end{equation}

\begin{equation}
f_t= \begin{cases}
    \frac{f_{t-i}+N\cdot f_{avg}}{N+1} &  f_{avg} - f_{var} \leq f_i \leq f_{avg} + f_{var} \\
    f_i & f_i \leq f_{avg} - f_{var} \\
    f_{i} & f_i \geq f_{avg} + f_{var}\\
\end{cases}
\label{eq_11}
\end{equation}

\noindent Where \(f_{avg}\) is the average blendshape estimate summed over \(N\) previous individual states \(f_{t-i}\), with \(f_t\) being the current state being considered for smoothing, \(f_{var}\) is the variance between N blendshape states in terms of the average value, and \(f_{smooth}\) is the new smoothed blendshape state. The strength of the smoothing effect and resulting filter lag can be adjusted by changing the number of previous blendshape estimates used to calculate the average. This smoother is especially useful for blendshapes with large sudden dynamic movements, like Blink or JawOpen.

\begin{figure}
    \centering
    \includegraphics[width=\columnwidth]{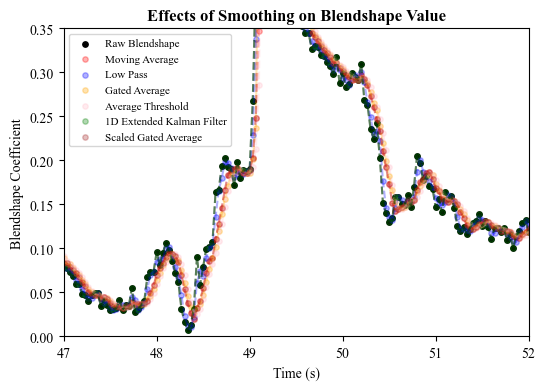}
    \caption{Demonstration of different smoothing algorithms on blendshape estimates over a much narrow time range.}
    \label{fig_smoothing}
\end{figure}

\section{Experiments and Results}

\subsection{Dataset and Implementation}

Two different data sources were used for training. 131 raw images taken with an iPhone 12 pro were used for rough segmentation and feature identification but not training. 
An unreal Engine 4 metahuman avatar that had been configured to receive LiveLink blendshape commands was used to create synthetic training data for each blendshape, usually 200-600 pictures at a resolution matching a standard webcam (400x600 pixels). Four different training steps were required, feature section/segmentation, regression, weighting function, and smoothing. Correlation and covariance analysis with other statistical methods was used to identify relevant landmarks for regression. Regression analysis is used to create a base regression model, optimized for accuracy, memory usage, and complexity. Nonlinear weighting functions are optimized based on comparison with a series of autocorrelation tests and video analysis. Finally smoothing was optimized to reduce inter-frame jitter.

\begin{figure*}[h!]
    \centering
    \includegraphics[width=0.85\textwidth]{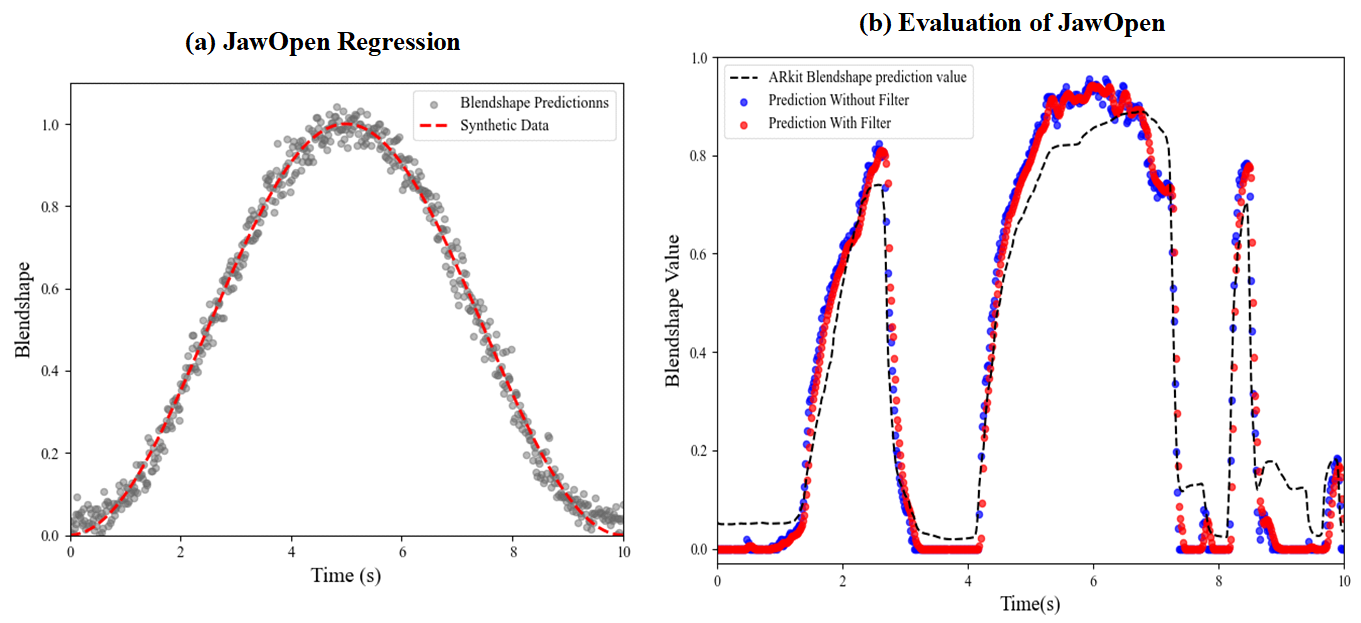}
    \caption{(a) Shows JawOpen blendshape predictions (gray dots) compared with synthetic test data (dashed red). (b) Demonstrates the high degree of similarity between Apple ARKit 6\cite{noauthor_arkit_nodate} (dashed line) and the new model without filtering (blue) and with filtering (red).}
    \label{fig_4v2}
\end{figure*}

Testing was conducted using 18209 frames taken from 21 videos captured using an iPhone 14 with a resolution of 1920x1440 pixels. ARKit has access to an IR camera and a LIDAR sensor as part of the sensor suite, and is used for algorithm performance comparison. Higher resolution was required because of ARKit. Six different testers, 4 male and 2 female, recorded in a variety of environments and lighting conditions. Six compound expressions and 9 types of head movements are included for each tester. Testers were trained to perform facial expressions using an example avatar for each specific expression. Testers were asked to mimic avatar expressions, and responses were recorded. Manual annotation was provided using a trained observed who was able to observe the same model avatar expressions.

\subsection{Regression Performance}

After verifying that landmark data was suitable for regression, we developed the complete system described in the methods section. An affine transformation was developed to transform the face into a common basis for regression analysis. Significant effort was spent identifying the most useful features for regression analysis and modeling. Next data was transformed to creates displacement vectors regression. Principal component analysis was used to first identify directions of maximum variance. Displacement was then estimated in terms of relevant anchor points, and converted into aspect ratios of the anchor points. Further data transformation was sometimes necessary for specific blendshapes. This was usually caused by heteroskedasticity or multicollinearity. Regression analysis was performed on the transformed landmark vectors for each blendshape. Each regression method has their own strengths and weaknesses, and has been chosen for maximum accuracy and performance. Heteroskedasticity and bias correction motivated the development of additional regression terms in our blendshape models. Adding these additional corrective terms restored dynamic range of blendshapes, allowing for full closure of the human eye under real world conditions. The exact regression models and weight functions used were independently optimized for each blendshape. 

During real world testing we observed significant bias and noise. This motivated the development of smoothing algorithms. We quantified the effects of these smoothing algorithms using statistical testing. BlinkLeft using a common dataset of 3200 detections from real world testing. As an example, the gated average smoother has the largest first order autocorrelation using 5 previous states instead of 10. The gated average smoother with 5 states imposes an average 59 millisecond delay before a blendshape prediction with an identical value is observed. This is a much smaller time delay compared to a standard interpolation algorithm. The memory used by each smoother is very small and is therefore suitable for low power computing applications. 

We observed close agreement between regression models and the training data. Further testing shown in \cref{fig_4v2}(a), where an Unreal Engine 5 human avatar model is controlled using LiveLink blendshape commands using a sine wave generating function shown with the red dashed line. We then used the complete regression system to estimate blendshape values, shown with the gray dots. The likely source of discrepancy between the source data and the estimates is the regression error introduced by the Mediapipe Holisitic 0.9.0.1 landmark regression process, especially the Blazeface\cite{bazarevsky_blazeface_2019} segmentation step. The regression model accurately matches the generating function, even at the limit of the dynamic range where the landmark detection model shows significant instability. 

We performed further statistical analysis to compare our blendshape estimation method with Apple ARKit 6\cite{noauthor_arkit_nodate}. First is a direct comparison, for a single blendshape using a reference video clip, demonstrated in \cref{fig_4v2}(b). Estimated JawOpen from ARKit and our regression model both with and without the gated average smoothing algorithm. There is a high degree of consistency between our algorithm and the ARKit response, with a mean deviation of -0.0005 a d mean squared deviation of 0.007 over the test video. We interpret regions with increased deviation at the end of a movement cycle as interpolation or smoothing algorithm artifacts built into the ARKit JawOpen model, and have verified that the mouth/jaw is closed at the specific times investigated. Our blendshape model shows consistently faster response at the beginning of a movement cycle, which we further interpret as the result of more responsive smoothing methods. There is noticeable jitter in estimated blendshape value at the upper limit of the dynamic range ($>0.8$) caused by instability in the Mediapipe Holistic 0.9.0.1 model at these times, partly compensated by the gated average smoother. Our model is capable of detecting small, real time changes in micro-expression that does not seem to be observable with ARKit.

\subsection{Comparative Analysis}

\begin{table*}[ht!]
\caption{Comparison Between ARKit and our methods, and response similarity between methods}
\resizebox{\textwidth}{!}{%
\begin{tabular}{l|ccc|ccc|cccccc}
    \hline
    \multirow{2}{*}{\textbf{Blendshape}} & \multicolumn{3}{c|}{\textbf{ARKit}\cite{noauthor_arkit_nodate}} & \multicolumn{3}{c|}{\textbf{Our Model}} & \multicolumn{6}{c}{\textbf{Cross-Comparison}} \\
     & Pr & Re & F1\cite{taha_metrics_2015} & Pr & Re & F1\cite{taha_metrics_2015} & P Corr\cite{pearson_liii_1901} & S Corr\cite{spearman_correlation_1910} & $\xi$\cite{chatterjee_new_2021} & Accuracy & MSD & Deviation\\
    \hline
    EyeBlinkLeft & 1.00 & 1.00 & 1.00 & 1.00 & 1.00 & 1.00 & 0.576 & 0.572 & 0.305 & 0.752 & 0.061 & 0.216 \\
    EyeLookDownLeft & 0.80 & 1.00 & 0.88 & 0.61 & 1.00 & 0.76 & 0.657 & 0.678 & 0.393 & 0.876 & 0.015 & 0.074 \\
    EyeLookInLeft & 0.80 & 1.00 & 0.88 & 0.50 & 1.00 & 0.66 & 0.489 & 0.504 & 0.402 & 0.773 & 0.051 & 0.224 \\
    EyeBlinkRight & 1.00 & 1.00 & 1.00 & 1.00 & 1.00 & 1.00 & 0.674 & 0.675 & 0.410 & 0.794 & 0.042 & 0.282 \\
    EyeLookDownRight & 0.80 & 1.00 & 0.88 & 0.61 & 1.00 & 0.76 & 0.658 & 0.679 & 0.385 & 0.876 & 0.015 & 0.074 \\
    EyeLookInRIght & 0.83 & 1.00 & 0.90 & 0.80 & 0.80 & 0.80 & 0.567 & 0.606 & 0.400 & 0.914 & 0.007 & 0.133 \\
    JawOpen & 1.00 & 1.00 & 1.00 & 1.00 & 1.00 & 1.00 & 0.921 & 0.886 & 0.682 & 0.965 & 0.001 & 0.039 \\
    MouthRight & 0.88 & 1.00 & 0.94 & 0.80 & 1.00 & 0.88 & 0.636 & 0.537 & 0.349 & 0.961 & 0.001 & 0.025\\
    MouthLeft & 0.88 & 1.00 & 0.94 & 1.00 & 0.62 & 0.76 & 0.610 & 0.616 & 0.425 & 0.945 & 0.002 & 0.042\\
    MouthSmileLeft & 0.88 & 1.00 & 0.94 & 1.00 & 0.62 & 0.76 & 0.932 & 0.953 & 0.740 & 0.867 & 0.017 & 0.073 \\
    MouthSmileRight & 0.88 & 1.00 & 0.94 & 0.80 & 1.00 & 0.88 & 0.907 & 0.943 & 0.702 & 0.861 & 0.019 & 0.067 \\
    MouthUpperUpLeft & 0.88 & 1.00 & 0.94 & 1.00 & 0.62 & 0.76 & 0.474 & 0.467 & 0.407 & 0.925 & 0.005 & 0.078 \\
   MouthUpperUpRight & 0.88 & 1.00 & 0.94 & 0.80 & 1.00 & 0.88 & 0.496 & 0.504 & 0.387 & 0.927 & 0.005 & 0.080 \\
   BrowDownLeft & 0.77 & 1.00 & 0.87 & 0.63 & 1.00 & 0.77 & 0.699 & 0.544 & 0.404 & 0.698 & 0.091 & 0.337 \\
   BrowDownRIght & 0.77 & 1.00 & 0.87 & 0.75 & 0.85 & 0.80 & 0.717 & 0.518 & 0.373 & 0.756 & 0.059 & 0.235 \\
  Average & 0.87 & 1.00 & 0.93 & 0.82 & 0.90 & 0.83 & 0.667 & 0.645 & 0.451 & 0.859 & 0.019 & 0.132 \\
  \hline
\end{tabular}
}
\label{tab_comp}
\end{table*}

Next we wanted to compare model performance with an existing system, specifically ARKit. This is complicated by a lack of objective truth data and full model details not being publicly available. Blendshape expression definitions are tied to specific mesh movements, and interaction between different blendshapes is used to create more realistic faces. We are limited to the set of expressions using arbitrarily defined blendshapes, which may not correspond with physical face movements. We therefore devised several statistical methods to compare response accuracy. These methods include different types of correlation analysis and a modified $F1$ score for specific detection events. Results are shown in \cref{tab_comp}.

Our correlation analysis focused on the response similarity between ARKit 6 and our blendshape regression approach. We used several different types of correlation, including Pearson product correlation\cite{pearson_liii_1901}, Spearman's rank correlation\cite{spearman_correlation_1910,spearman_proof_1904}, and $\xi$ correlation\cite{chatterjee_new_2021}. The usage of multiple measures of correlation is effective to enhance the credibility of our experimental results. Specifically, Pearson correlation measures the strength of the linear relationship between variables while Spearman and $\xi$ correlation measure the strength of a monotonic relationship which is not necessarily linear. Spearman and $\xi$ are also less sensitive to outliers because of extra data preprocessing.

For a more objective comparison we used a modified $F1$ scoring criteria based on test dataset annotations. These annotations recorded when an expression could be observed to occur, but not specific expression intensity. The calculation of $F1$-score is described as:

\begin{equation}
    F1=\frac{(2\times P \times R)}{(P+R)}
    \label{eq_12}
\end{equation}

\noindent where P is precision and R represents recall. Precision measures the ability of preventing the prediction of false positive and recall represents the probability of missing positive values. Precision and recall are determined by:

\begin{equation}
    Precision=\frac{n_{TP}}{n_{TP}+n_{FP}}
    \label{eq_13}
\end{equation}

\begin{equation}
    Recall=\frac{n_{TP}}{n_{TP}+n_{FN}}
    \label{eq_14}
\end{equation}

\noindent Where $n_{TP}$ is the number of true positive (TP) observations, $n_{FP}$ the number false positive (FP) observations and $n_{FN}$ false negative (FN) observations. The decision criteria for each distinct expression event is determined by containment within sensitivity and time thresholds defined for an expression event for each blendshape:

\begin{equation}
    \delta_{f_b,t_k} = \begin{cases}
    1\; \text{if} \;f_{min}\leq f_{b};\; t_{min} \leq t_{k} \leq t_{max}\\
    0\; \text{if} \;f_{min}\geq f_{b};\; t_{k} < t_{min}; t_{k} \geq t_{max}\\
    \end{cases}
\end{equation}

\begin{equation}
     n_{TP} = \sum_{t=0}^{t_N}{\delta_{f_b,t_k}}
\end{equation}

\begin{equation}
    n_{FN} = n_{algo} - n_{TP}
\end{equation}

\begin{equation}
    n_{FN} = n_{ref} - n_{TP}
\end{equation}

\noindent where $\delta$ is the determination of each detected expression events, $f_{min}$ is a minimum sensitivity threshold optimized based on human observer decisions, $t_{min}$ and $t_{max}$ define a time window around unique expression events, $t_k$ representing a specific video frame number, $f_b$ blendshape value. The sum of decisions on detected expression events determines the number of TP, FP, and FN observations. $n_{ref}$ refers to the total number of annotated expression events and $n_{algo}$ is the number of correct algorithm decisions. The time window defined by:

\begin{equation}
    t_{min} = t_{anno} - \Delta_{k}
\end{equation}

\begin{equation}
    t_{max} = t_{anno} + \Delta_{k}
\end{equation}

where $t_{anno}$ is the frame number of a specific expression event, and $\Delta_{k}$ is an adjustable accuracy threshold. As an example this accuracy threshold is approximately 15 frames, but needs to be adjusted for videos with faster frame rates.

\begin{figure}[h!]
    \centering
    \includegraphics[width=\columnwidth]{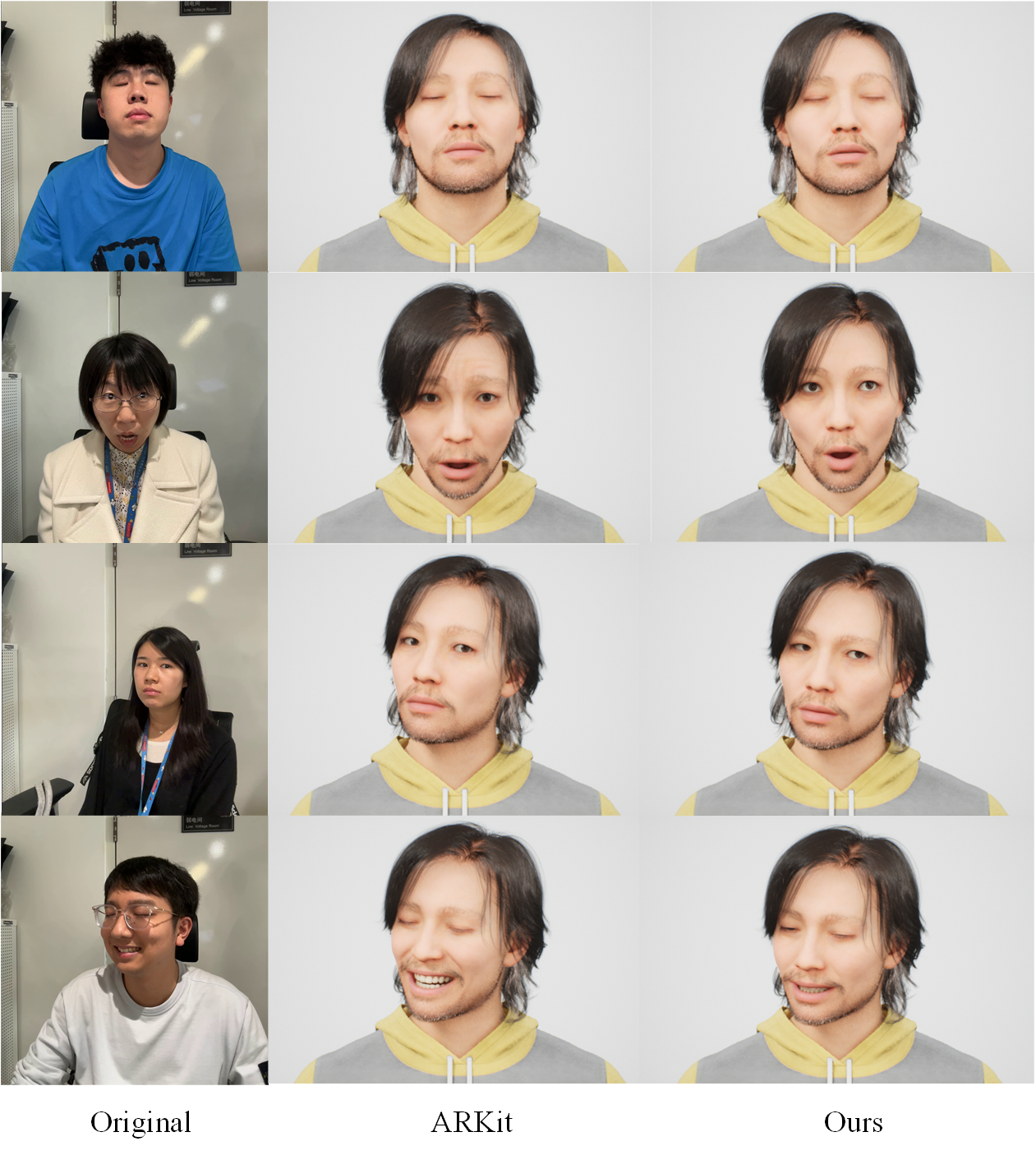}
    \caption{Response comparison between raw image (left column), ARKit (middle coloumn) and our algorithm (right column)}
    \label{fig_vis}
\end{figure}

\begin{figure*}[t!]
    \centering
    \includegraphics[width=0.8\textwidth]{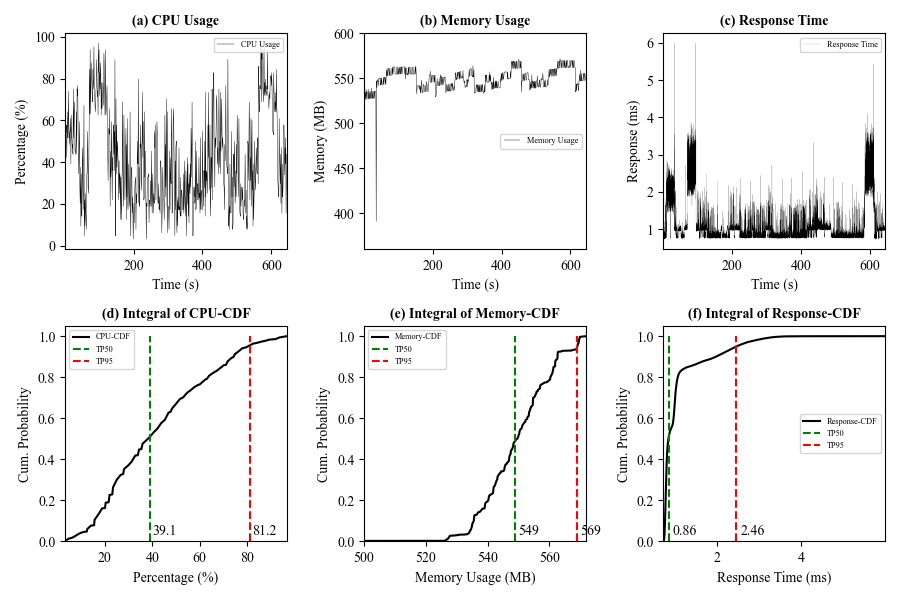}
    \caption{Performance validation data of 3D avatar algorithm. (a), (b), (c) The CPU usage/memory usage/response time over time. (d), (e), (f) CDF graph of (a), (b), (c), which shows the location of TP50/TP95 value.}
    \label{fig_5}
\end{figure*}

We display the correlation between our algorithm and ARKit as well as accuracy metrics of F1-score in \cref{tab_comp}. From the result we can see that most of our blendshapes calculation are correlated well with ARKit while gaining acceptable performances of subjective observation. For EyeBlinkLeft, EyeBlinkRight and JawOpen, we achieve the 100\% accuracy score of precision, recall and $F1$-score, which indicate the same subjective impression as ARKit. However, the estimation of other left side’s blendshapes are not as accurate as right side which can be a direction for further optimization. As for the overall F1-score, we reach to a small difference (9.53\%) comparing to ARKit, which demonstrates that the usage in real world is also acceptable. 

Visual response comparison is shown in \cref{fig_vis}. This highlights the strong response similarity between our algorithm and real human face expressions. We demonstrate slight improvement in response accuracy for eye and mouth blendshapes under testing conditions compared to ARKit 6. ARkit shows significant robustness, and further work is needed to bring our system to the same standard.

\subsection{Hardware Performance}
 
A hardware performance analysis was conducted to assess the suitability of our calculation process for real world application. Testing was done using a 12th Gen Intel(R) Core(TM) i7-12700H CPU with 14 cores. We measured the CPU, memory usage and response time for the entire test dataset. All 21 test videos are inputted to system and processed sequentially with performance data simultaneously recorded, shown in \cref{fig_5}(a)-(c). Statistical analysis was conducted on the performance measurements, shown with cumulative probability plots (d)-(f). The 50th and 95th percentile of CPU usage on 1 core range between 39.1\% and 81.2\%, while total memory usage ranges between 549MB and 569MB during indicates a stable working state. More than 90\% of the utilized memory is consumed by the MediaPipe Holistic model. All processing steps including the landmark detection process are executed on the CPU. Algorithm response time 50th and 95th percentile values range between 0.86ms and 2.46ms. Two versions of our algorithm were developed, using either python or C++. The python version tested during this experiment, and is able to perform real time blendshape calculation, suitable for smartphone and tablet applications.

\section{Conclusion}

We were able to successfully develop a statistical model to convert landmark estimates to blendshapes for 3D human avatar animation. This involved affine transformation, segmentation, data transformation, autocorrelative regression, and smoothing. Development of each process was supported using extensive statistical analysis. We were able to reproduce blendshape response behavior used by unreal engine 4 metahuman avatars and apple ARKit 6, and produce comparable overall accuracy. We were able to achieve an average mean squared deviation of 0.019 from ARKit 6. Our system does not require significant training data to obtain optimal performance. This algorithm provides improved VR capability for low power computing devices that lack GPU or similar capability and limited sensor capability, using a standard webcam instead of the integrated LIDAR-IR-depth camera system used by iPhones. Our algorithm is relatively hardware independent, adding improved VR capability to Windows and Linux systems where ARKit cannot be deployed. 

Future work could focus on developing optimized weighting functions, replacing the MediaPipe Holistic Model with our own landmark regression method. Optimizing weight functions for regression could significantly optimize the results from our regression models, as well as partially correcting for data limitations. Replacing the MediaPipe holistic model with a more robust keypoint detection method could remove some of the model deformation effects we observed during development. This could significantly improve the robustness of our blendshape estimation model under real world conditions. Overall, we were able to successfully develop a useful blendshape conversion algorithm using a relatively small training dataset. 


\bibliographystyle{abbrv-doi}

\bibliography{Avatarpaper}
\end{document}